\documentclass[aps,prl,twocolumn,showpacs,superscriptaddress]{revtex4}
\usepackage{amssymb}
\usepackage{graphicx}
\usepackage{dcolumn}
\usepackage{bm}
\usepackage{amsmath}

\begin{document}

\title{ Precise Control of Molecular Dynamics with a Femtosecond
Frequency Comb - A Weak Field Route to Strong Field Coherent
Control}

\author{Avi Pe'er}
\affiliation{JILA, National Institute of Standards and Technology
and University of Colorado, Boulder, CO 80309-0440}
\author{Evgeny A. Shapiro}
\affiliation{Department of Chemistry, University of British
Columbia, Vancouver, Canada}
\author{Matthew C. Stowe}
\affiliation{JILA, National Institute of Standards and Technology
and University of Colorado, Boulder, CO 80309-0440}
\author{Moshe Shapiro}
\affiliation{Department of Chemistry, University of British
Columbia, Vancouver, Canada}

\affiliation{Department of Chemical Physics, Weizmann Institute of
Science, Rehovot 76100, Israel}
\author{Jun Ye}
\affiliation{JILA, National Institute of Standards and Technology
and University of Colorado, Boulder, CO 80309-0440}

\begin{abstract}

We present a general and highly efficient scheme for performing
narrow-band Raman transitions between molecular vibrational levels
using a coherent train of weak pump-dump pairs of shaped ultrashort
pulses. The use of weak pulses permits an analytic description
within the framework of coherent control in the perturbative regime,
while coherent accumulation of many pulse pairs enables near unity
transfer efficiency with a high spectral selectivity, thus forming a
powerful combination of pump-dump control schemes and the precision
of the frequency comb. The concept is presented analytically and its
feasibility and robustness are verified by simulations of dynamics
in Rb$_2$. We consider application of this concept to the formation
of stable, deeply bound, ultracold molecules.
\end{abstract}


\maketitle

Although mode-locked lasers emit broadband ultrashort pulses, they
can be utilized to perform frequency selective excitation just like
narrow-band CW lasers due to their precise frequency comb
\cite{FrequencyComb}. This spectral selectivity is explained by the
very long inter-pulse phase coherence, which allows for coherent
accumulation of the excitation amplitudes from multiple pulses in an
excited material system, similar to a generalized Ramsey experiment.
This idea led to the realization of direct frequency comb
spectroscopy in atomic systems \cite{DFCS}. Here we propose to apply
the principle of coherent accumulation, combined with weak field
coherent control, to precisely control molecular dynamics at high
efficiencies.

While analysis of coherent quantum control is relatively simple in
the weak field perturbative domain \cite{WeakFieldCoherentControl},
extension to strong fields is not straightforward. Analytic models
exist only for simple cases \cite{StrongFieldControl,
Vitanov@Bergmann_STIRAP_AAMOP_2001} and solutions are often found by
numerical optimizations \cite{Rabitz@Kompa_Science_2000}. The core
of our approach is to exploit analytic perturbative models to design
``ideal'' weak pulses and to achieve the high overall efficiency by
coherently accumulating many such pulses. This avoids the
complication of strong field design while gaining high spectral
selectivity offered by the frequency comb. Since the maximal number
of accumulated pulses is inherently limited by the coherence time of
the material ensemble, we expect our approach to be applicable
particularly well to ultracold atomic / molecular ensembles, where
coherence times are long.

\begin{figure}[tb]
\begin{center}
\includegraphics[width=8.6cm] {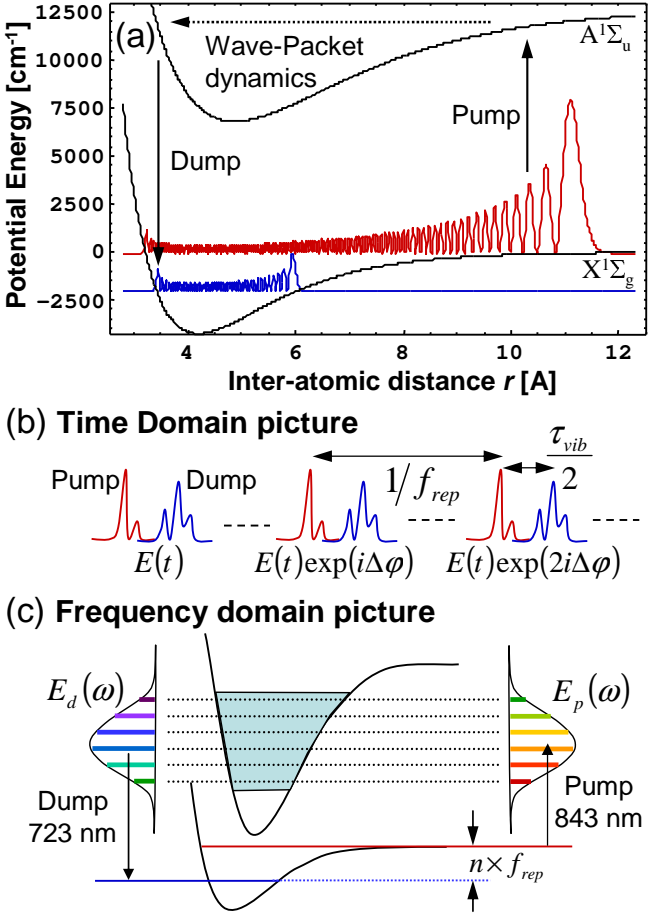}
\caption{\label{Fig1-concept} Basic Raman control scheme. (a)
Typical electronic potentials and vibrational levels (here, Morse
potential fits of Rb$_2$). Population is transferred from the input
vibrational level near the dissociation limit ($v'=130$) to a deeply
bound level ($v=45$), mediated via a broadband wavepacket in the
excited electronic potential. (b) Time domain picture. A train of
phase coherent pump-dump pulse pairs interacts with the molecule.
Pulse pairs are shaped to achieve efficient population transfer. The
intra-pair time is half the vibration time ($\tau_{vib}$) of the
intermediate wavepacket and the inter-pair time is the repetition
time of the source $1/f_{rep}$. (c) Frequency domain picture. A
tooth-to-tooth match between the pump and the dump frequency combs,
with $f_{rep}$ equal to a sub-harmonic of the net Raman energy
difference, locks the relative phase between pulse pairs to the free
evolving Raman phase.}
\end{center}
\end{figure}

Motivated by the goal to produce deeply bound ultracold polar
molecules from loosely bound Feshbach molecules, we consider the
Raman transition shown in Fig. \ref{Fig1-concept} from a single
vibrational level embedded in a dense environment of other levels
near the dissociation limit, to a single deeply bound vibrational
level. Loosely bound Feshbach molecules can be generated with high
efficiency via adiabatic sweeping through a magnetic Feshbach
scattering resonance in an ultracold atomic ensemble
\cite{FeshbachMolecules}. Since molecules are not amenable to
standard laser cooling techniques, magneto-photo-association of cold
atoms is a major avenue pursued for obtaining cold molecules, which
now represents one of the most exciting research fields in cold
matter. So far, stable, ultracold ground state molecules have not
been produced in high densities \cite{GroundStateMolecules}, mainly
because of the poor wavefunction overlap between the spatially
spread continuum states of colliding atoms and the localized
molecular states. Feshbach molecules appear therefore to be an
important mid-stage towards deeply bound ultracold molecules.

When the energy level spacing near the input state is small compared
to the natural line-width of excited states, the application of CW
techniques is not straightforward. Any scheme relying on populating
an excited state during the transition is inherently limited because
most of the population will be lost to spontaneous decay during the
time needed to resolve the input state. Short pump-dump pulses could
eliminate spontaneous emission losses \cite{PumpDumpPA}, but then
the total interaction time is too short. A similar problem is met by
stimulated Raman adiabatic passage (STIRAP). For the adiabatic
passage to hold, the Rabi frequency $\Omega$ must be larger than the
decay rate $\Gamma$ from the intermediate state: $\Omega \gtrsim
\Gamma$ \cite{Vitanov@Bergmann_STIRAP_AAMOP_2001}. Since both the
power broadened pump line-width and the two-photon line-width of
STIRAP are of order $\Omega$, the possibility of population transfer
to neighboring levels should be considered. Thus, predictions for
using STIRAP to photo-associate an atomic Bose-Einstein condensate
depend strongly on modeling of the decay process and the atomic
continuum states \cite{BEC-STIRAP}.

To overcome these difficulties, we employ the scheme illustrated in
Fig. \ref{Fig1-concept}, which is based on a phase-coherent train of
shaped pump-dump pulse pairs. Each pulse pair is weak, i.e., it
transfers only a small fraction of the input population to the
target state. Coherent accumulation then enables a high overall
transfer efficiency. In the time domain, each pump pulse excites a
wave packet that starts to oscillate in the excited electronic
potential. After half a vibration, this wave-packet reaches the
inner turning point where the dump pulse drives it to the target
state. Since population appears on the excited potential only for
half a vibration, this scheme eliminates spontaneous emission
losses. In between pulse-pairs the system is left in a coherent
superposition of the input and target states that evolves freely.
Therefore, in order to enable coherent accumulation of population at
the target state, the temporal phase difference between pulse pairs
$\Delta\varphi$ should match the phase of the free evolving Raman
coherence. In the frequency domain, the combs of the pump and the
dump pulses must overlap tooth to tooth and the repetition rate
$f_{rep}$ must match a sub-harmonic of the Raman energy difference.

Let us clarify the role of the different amplitudes and phases
involved. All pulse pairs share a common temporal (spectral) shape
and phase ($E(t)$ or $E(\omega)$), which is designed to maximize the
transfer efficiency for a single pulse pair. The relative phase
between successive pulse pairs, $\Delta\varphi$, is then controlled
via stabilization of the frequency comb to achieve coherent
accumulation.

For the very first dump pulse to drive all the excited population to
the $empty$ target state it's ``pulse area'' should be $\sim$$\pi$.
After the second pump pulse however, the excited population is about
equal to the population already in the target state, so now only a
$\pi/2$ ``area'' dump pulse of the appropriate phase is required to
perform the transfer, just like in a Ramsey experiment. In general,
the fraction $p$ of population excited (or dumped) is related to the
``pulse area'' $A$ by $\sin^{2}\left(A/2\right)=p$. As population is
accumulated in the target state and depleted from the input state,
the dump ``area'' of the $n$th pulse $A_{d}[n]$ should decrease and
the pump ``area'' $A_{p}[n]$ increase for each pulse pair to
transfer the same fraction of population (for $N$ pulses),
\begin{eqnarray}
\label{CA_eq1} \sin^{2}\left(\frac{A_{d}[n]}{2}\right)=\frac{1}{n},\
\ \ \ \sin^{2}\left(\frac{A_{p}[n]}{2}\right)=\frac{1}{N-n+1}.
\end{eqnarray}

The ``pulse area'' is not a well defined quantity outside the
context of a two-level system. However, the ratio of populations in
the excited wave-packet and the input (target) state defines an
effective area for the pump (dump) pulses. This concept proves
useful mainly for weak pulses, where the excitation is predominantly
a one-photon process. In general, for any given pump series, a dump
series can be matched according to the $n$th fraction of excited
population $p[n]$. Clearly the very first dump pulses and the last
pump pulses are of areas near $\pi$ and cannot be considered weak,
but for a large $N$, the majority of the population is transferred
by the accumulative effect of all pulses, which are mostly weak.

For an efficient pump-dump process it is required that the
wave-packet $|\psi_p\rangle$, excited by the pump from the input
state $|i\rangle$ and propagated for half a vibration, will overlap
perfectly with the wave packet $|\psi^{r}_{d}\rangle$ that would
have been excited from the target state $|t\rangle$, by the time
reversed dump. For weak pulses we can express these two
wave-packets, using first order perturbation theory, as
\begin{eqnarray}
\label{CA_eq3}
|\psi_p\rangle=\sum_{\omega}
E_{p}(\omega)\exp\left[i\phi_{D}(\omega)\right]|\omega\rangle\langle\omega|
d_{el}|i\rangle \nonumber \\
|\psi^{r}_{d}\rangle=\sum_{\omega}
E^{r}_{d}(\omega)|\omega\rangle\langle\omega|d_{el}|t\rangle,\
\ \ \ \ \ \ \ \ \ \ \ \ \ \ \
\end{eqnarray}
where $|\omega\rangle$ denotes the vibrational states in the excited
potential using the detuning $\omega$ from the pulse carrier
frequency as a vibrational index,
$F_{p}(\omega)=\langle\omega|d_{el}|i\rangle$ and
$F_{d}(\omega)=\langle\omega|d_{el}|t\rangle$ are the pump and the
dump transition dipole matrix elements, which under the Condon
approximation are propotional to the Franck-Condon factors
$\langle\omega|i\rangle,\langle\omega|t\rangle$.  $d_{el}$ is the
electronic transition dipole moment and $E_{p}(E^r_d)$ is the
spectral amplitude of the pump (time-reversed dump) field.
$\phi_{D}(\omega)$ is the spectral phase acquired by the wave-packet
between the pulses, which reflects both the delay of half a
vibration and the dispersion of the wave packet as it oscillates in
the anharmonic excited potential. For the Morse potential fit, used
later in the simulations, analytic expressions exist for the
vibrational states and energies \cite{MorseData}, so both the dipole
matrix elements and $\phi_{D}(\omega)$ are known.

\begin{figure}[tb]
\begin{center}
\includegraphics[width=8.6cm] {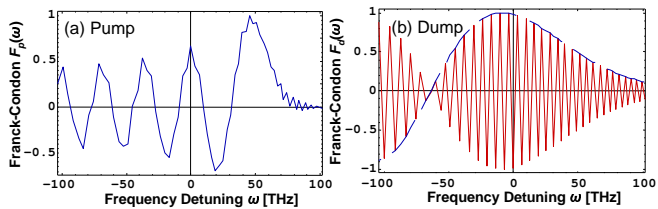}
\caption{\label{Fig2-FCs} Franck-Condon factors ($\langle v|v'\rangle$) vs.
frequency detuning from the central state of the wavepacket; (a) for
excitation from level $v'=130$ in the ground electronic potential to
levels centered around $v=162$ in the excited potential and (b) for
de-excitation of these levels to $v=45$ in the ground potential.}
\end{center}
\end{figure}

As a result, perfect overlap of the two wave packets can be achieved
by shaping the pump field according to the dump dipole matrix
elements and vice versa:
\begin{eqnarray}
\label{CA_eq4} E_{p}\left(\omega\right)\propto
F_{d}\left(\omega\right)A\left(\omega\right) \
\ \ \ \ \ \ \ \ \ \ \ \ \ \ \ \ \ \nonumber \\
E_{d}\left(\omega\right)\propto
F_{p}\left(\omega\right)A\left(\omega\right)
\exp\left[i\phi_{D}\left(\omega\right)\right],
\end{eqnarray}
where $A$ is an arbitrary spectral amplitude, common to both fields.
Intuitively, this spectral shaping avoids pumping of what cannot be
dumped (due to a node in the dump dipole matrix elements), and vice
versa. An example of pump (dump) Franck-Condon factors is shown in
Fig. \ref{Fig2-FCs}a(b). The fast oscillation (alternating sign) of
the Franck-Condon factors for the dump pulse is canceled by the
relative delay of half a vibration between the two pulses, so the
pump should only be spectrally shaped according to the slow envelope
of $F_d(\omega)$. Such shaping is easily achieved with current
ultrafast pulse shapers \cite{Weiner_RSI_2000}. It may seem
surprising that a broadband pulse can dump a broadband wave-packet
to a single state, but since we tailored the wave-packet to the
pulse, such that all the transition amplitudes from levels within
the wave-packet interfere constructively only at the target state,
this is no contradiction. While the shaped pump-dump pulse pairs
achieve spectral selection of a single vibrational level, coherent
accumulation is key to further refine the spectral resolution to
address rotation and hyperfine levels, as well as to accomplish near
unity transfer efficiency.

To check the viability of our scheme, we numerically simulated the
molecular dynamics driven by a train of pulses as discussed. The
simulation is based on a split-operator code
\cite{Garraway@Suominen_RPP_1995} that solves the time-dependent
Schr\"odinger equation (within the rotating wave approximation) with
three wave-packets on three potential surfaces, coupled by two
arbitrary pulses. According to Eq. (\ref{CA_eq4}), the pulses can be
shaped in two stages: differential shaping that ensures overlap of
the wave-packets by matching the pulses to the different Franck
Condon spectral responses; and common shaping, which affects the
overall shape of both wave-packets (e.g. common chirping). First,
the desired effect of differential shaping was verified. Indeed,
with weak pulses ($\leq\pi/10$) shaped according to the Spectral
Franck-Condon function shown in Fig. \ref{Fig2-FCs}, the overlap of
the pumped and dumped wave-packets was practically unity ($>0.999$).
We then explored the effect of common shaping on the pump-dump
process. Within the perturbative discussion relevant to Eq.
(\ref{CA_eq4}), the common spectral amplitude $A\left(\omega\right)$
is completely arbitrary, and for weak enough pulses, this is
verified by our simulation. Yet, it is desirable to minimize the
total number of pulses for both practical reasons (more tolerant
phase locks) and fundamental ones (the total interaction time is
limited by the coherence time of the input state), so how strong can
the pulses be and still qualify as ``weak''? The limiting power
level is where two-photon (Raman) processes by one pump pulse become
pronounced, and here spectral shaping will have an important effect.

In many cases of molecular dynamics, positively chirped (red to
blue) excitation pulses can strongly suppress Raman processes that
adversely affect the input wave-packet during the pulse
\cite{PositiveChirp}, leaving the excitation, although strong,
essentially one-photon. The reason is that within the Franck-Condon
window, the excited potential is usually steeper than the ground
potential (e.g. excitation from the ground vibrational level,
localized at the zero slope of the ground potential), so population
initially excited by the red part of the pulse cannot be later
de-excited by the blue part because there are no available levels to
de-excite to. It is clearly shown in simulation that when the pulses
are positively chirped to be longer than the vibration time of both
the input and the target states, over 50\% of population can be
selectively transferred between two deeply bound vibrational levels
with one pump-dump pair. Thus, chirping the pulse can improve
considerably the dumping efficiency for the first dump pulse that is
necessarily strong because it dumps to an empty target state.

The common chirping can help resolve vibrational structure around
the target state deep in the molecular well, yet it cannot resolve
rotational / hyperfine structure, and certainly not the dense
environment around the input state. Here, the combination of
coherent control and coherent accumulation proves powerful -
coherent control techniques (pulse shaping, chirping) are used to
achieve a precise state match between specific initial and target
states, while coherent accumulation allows high spectral selectivity
and total transfer efficiency.

\begin{figure}[tb]
\begin{center}
\includegraphics[width=8.6cm] {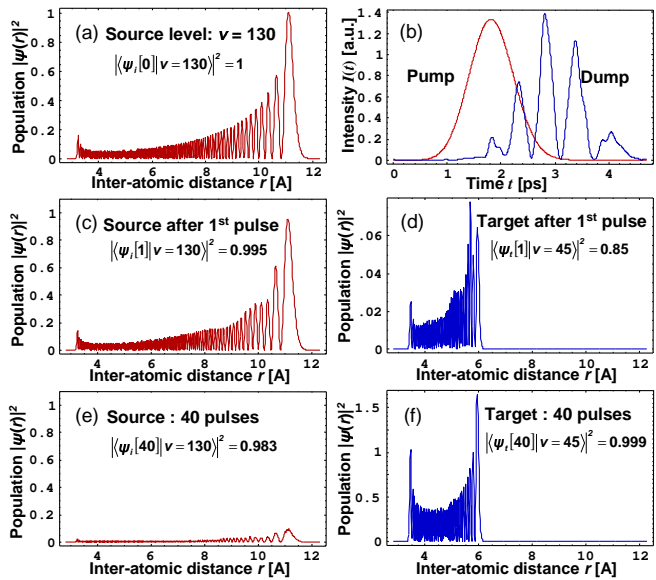}
\caption{\label{Fig3-simulation} Simulation results for the coherent
accumulation process. (a) the input state population density, (b)
the intensity temporal profile of the pulses (since the pulses are
strongly chirped it also represents the shaped power spectrum). (c)
and (d) are the input and target wave-packet population densities
after one pulse-pair, and (e) and (f) are the corresponding results
after 40 pulses. The energy density was $4.3x10^{-5}$ j/cm$^2$ per
pulse for the pump, and varied from $1.7x10^{-2}$ to $5x10^{-5}$
j/cm$^2$ for the dump}
\end{center}
\end{figure}

We simulated the full coherent accumulation process in various
scenarios. Figure \ref{Fig3-simulation} shows simulation results for
the interaction of a train of 40 pulse-pairs with the molecule at a
repetition time of 10 ns, assuming a 30 ns radiative lifetime for
the excited states. The pump pulse area was fixed to $~\pi/6.6$,
exciting about 5.7\% of the input population each time, and the dump
area was varied to match with the accumulation progress. The pulses
were $\sim10$ nm in bandwidth (100 fs transform limited), chirped
out to $\sim1.5$ ps by dispersion of 500,000 fs$^{2}$. The pump
pulses depleted $>$90\% of the input population, and when the
inter-pulse phase was tuned to the Raman condition, 95\% of this
population reached the target state. The purifying nature of the
coherent accumulation process is demonstrated by the obtained final
wave-packet - practically a single state.

The coherent accumulation process proves to be quite robust against
intensity fluctuations. Simulations show that scaling the intensity
of the pump or the dump pulse train (or both) by a factor of two
leaves the total transfer efficiency constant within a few percent.
The exact variation of dump pulse area according to the accumulated
population (Eq.(1)) is also not critical. Even if the dump area is
kept constant, the transfer efficiency is $>$50\% over a range of
factor of two in intensity.

Due to the high density of levels near the input state, it is
inherently impossible to avoid leakage of population to nearby
levels through two-photon Raman processes, which is exactly why CW
techniques, such as STIRAP, require caution. Although our comb
scheme is no exception, the deleterious effects of this leakage are
suppressed for two reasons. First, for a single pulse-pair the
leakage is diminished by the use of weak, mainly ``one photon''
pulses. Second, assuming the comb is not matched to the energy
spacing of nearby levels, after a large number of pulses $N$ the
leakage process resembles an incoherent random walk, thus scaling as
$\sqrt{N}$, whereas the coherent depletion of the input state scales
linearly as $N$, causing it to dominate. In our simulation, although
the input is depleted by $90\%$ after 40 pulses, it remains $>$98\%
pure.

Although we considered wave-packet dynamics in one excited
potential, the scheme is easily generalized to dynamics in a set of
coupled potentials (e.g spin-orbit). The pulses should then be
designed according to the (more complex) shape and phase of the
transition dipole matrix elements to the set of excited potentials.
In addition, since ultrashort pulses are used, the excited
wave-packets are broad and deeply bound within the excited
potential, so their dynamics is fast. Consequently, the scheme is
immune to small perturbations, such as hyperfine interactions, that
affect the inter-atomic potential near dissociation, and are usually
not well known.

To conclude, the presented scheme is a unique and powerful
combination of frequency domain control (comb) and time-domain
control (molecular dynamics). As such, it enables performance of
coherent control tasks with both high efficiency and unprecedented
spectral resolution. Due to the use of weak pulses the process can
be analyzed within a perturbative model, thus opening an analytic
path to strong field problems that were so far accessible only by
numerical optimizations. We believe the scheme is very
general and will find applications in many areas of coherent
control. Specifically, it can be used to produce deeply bound
ultracold molecules.

We thank P. Julienne, P. Zoller, and D. Jin for discussions. Work at
JILA is funded by NSF, DOE, and NIST. A. Pe'er thanks the Fulbright
Foundation for support.

\end{document}